\documentclass[preprintnumbers,prd,twocolumn,showpacs,floatfix,
nofootinbib,superscriptaddress]{revtex4}
\usepackage{graphicx}
\usepackage{epsfig}
\usepackage{bm}
\usepackage{amssymb}
\usepackage{float}
\usepackage{amsmath}
\usepackage{dcolumn}
\usepackage{cancel}
\usepackage[colorlinks]{hyperref}
\usepackage[usenames,dvipsnames]{color}
\hypersetup{
     breaklinks=true,
    pdfstartview={FitH},    
    colorlinks=true,       
    linkcolor=blue,          
    citecolor=red,        
    filecolor=magenta,      
    urlcolor=blue,           
    anchorcolor=green,      
    linktocpage=true
}

\newcommand{\Mpl}{M_{\textrm{Pl}}}
\renewcommand{\(}{\left(}
\renewcommand{\)}{\right)}

\def\al{\alpha}
\def\bet{\beta}

\def\Om{\Omega}
\def\sig{\sigma}
\def\lam{\lambda}
\def\ep{\epsilon}
\def\h{\mathcal{H}}

\def\N{\mathcal{N}}

\def\del{\delta}

\def\doi{http://doi.org}
\def\arxiv{http://arxiv.org/abs}
\def\t{\tilde}
\def\e{\mathrm{e}}
\def\r{\mathrm{r}}
\def\g{\mathrm{g}}
\def\h{\mathrm{h}}
\def\m{\mathrm{m}}
\def\rmt{\mathrm{t}}
\def\s{\mathrm{s}}

\begin{document}

\title{A class of quintessential inflation models with parameter space consistent with BICEP2}

\author{Md. Wali Hossain}
\email{wali@ctp-jamia.res.in}
\affiliation{Centre for Theoretical Physics, Jamia Millia Islamia,
New Delhi-110025, India}

\author{R. Myrzakulov}
\email{rmyrzakulov@gmail.com}
\affiliation{ Eurasian  International Center for Theoretical
Physics, Eurasian National University, Astana 010008, Kazakhstan}

\author{M.~Sami}
\email{sami@iucaa.ernet.in}
\affiliation{Centre for Theoretical
Physics, Jamia Millia Islamia, New Delhi-110025, India}

\author{Emmanuel N. Saridakis}
\email{Emmanuel\_Saridakis@baylor.edu}
 \affiliation{Physics
Division, National Technical University of Athens, 15780 Zografou
Campus,  Athens, Greece} \affiliation{Instituto de F\'{\i}sica,
Pontificia Universidad de Cat\'olica de Valpara\'{\i}so, Casilla
4950, Valpara\'{\i}so, Chile}

\begin{abstract}
In this paper, we focus on general features of quintessential
inflation which is an effort to unify inflation and dark energy
using a single scalar field. We describe a class of models of
quintessential inflation which can give rise to the tensor to scalar
ratio of perturbations consistent with recent BICEP2 measurements.
The scale of inflation in the model is around the GUT scale and
there is large parameter space consistent with the recent findings.
\end{abstract}

\pacs{98.80.-k, 98.80.Cq, 04.30.-w, 04.50.Kd}

\maketitle

\section{Introduction}

One of the  most outstanding and clean predictions of inflationary
paradigm is related to relic gravity waves \cite{Grishchuk:1974ny,Grishchuk:1977zz,
Starobinsky:1979ty,Allen:1987bk,Sahni:1990tx,Souradeep:1992sm,
Giovannini:1998bp,Giovannini:1999bh,Langlois:2000ns,Kobayashi:2003cn,
Hiramatsu:2003iz,Easther:2003re,Brustein:1995ah,Gasperini:1992dp,
Giovannini:1999qj,Giovannini:1997km,Gasperini:1992pa,Giovannini:2009kg,
Giovannini:2008zg,Giovannini:2010yy,Tashiro:2003qp,Sahni:2001qp} which are
 generated quantum mechanically in the early Universe. The primordial
tensor perturbations induce B mode polarization in microwave
background spectrum such that the effect depends upon the tensor to
scalar ratio of perturbations $r$. Since the effect was not observed
, the tensor to scalar ratio was supposed to be negligibly small.
However, the recent observations on CMB polarization has
demonstrated that the effect is sizeable, namely, the scalar to
tensor ratio of perturbations, $r=0.2_{-0.05}^{+0.07}$
\cite{Ade:2014xna} such that the scale of inflation is around the GUT
scale.

Quintessential inflation
\cite{Peebles:1999fz,Sahni:2001qp,Sami:2004xk,Copeland:2000hn,Huey:2001ae,
Majumdar:2001mm,Dimopoulos:2000md,Sami:2003my,Dimopoulos:2002hm,Rosenfeld:2005mt,
Giovannini:2003jw,Dimopoulos:2002ug,Nunes:2002wz,Dimopoulos:2001qu,
Dimopoulos:2001ix,Yahiro:2001uh,Kaganovich:2000fc,Peloso:1999dm,
Baccigalupi:1998mn,Hossain:2014xha}, a
 unified description of inflation and dark energy using a single
scalar field, is necessarily followed by kinetic regime responsible
for blue spectrum of relic gravity waves \cite{Giovannini:1998bp,
Giovannini:1999bh,Giovannini:1999qj,Sahni:2001qp}. These scenarios can
be classified into Type I and Type II. In first type, we consider
models for which the scalar field potential is exponentially steep
for most of the history of universe and only at late times the
potential turns shallow. In Type II, we place models with potentials,
shallow at early times followed by steep behavior thereafter. Ideally,
quintessential inflation requires a potential that could felicitate
slow roll in the early phase followed by approximately steep
exponential behavior such that the potential turns shallow only at late times. Steep
nature of potential is necessitated for the radiative regime to
commence and peculiar steep behavior is needed to realize the
scaling regime. However, the generic potentials
do not change their character
so frequently, they rather broadly come into two said categories:\\
Type A: The inverse power law and $\cos$ hyperbolic potentials belong to this
category. In this case one requires to assist slow roll by an extra
damping at early times. In Randall-Sundrum scenario \cite{Randall:1999ee,
Randall:1999vf}, the high energy corrections
to Einstein equations \cite{Shiromizu:1999wj} give rise to brane-damping which
assists slow roll along a steep potential \cite{Sahni:2001qp,Copeland:2000hn,Huey:2001ae,
Majumdar:2001mm,Maartens:1999hf}. As the field rolls down
its potential, high energy corrections cease leading to graceful
exit from inflation. Unfortunately, the tensor to scalar ratio in
this case is too large, $r\simeq 0.4$ to be consistent with
observations and the steep brane-world
inflation is therefore ruled out.\\
Type B: In this case, the field potential stays steep after
inflation. In this case, the late time behavior can be achieved by
invoking an extra feature in the potential. For instance, massive
neutrino matter with non-minimally coupled to scalar field can give
rise to minimum of the potential at late times when neutrinos become
non-relativistic \cite{Wetterich:2013jsa,Hossain:2014xha}.

 In this
paper, we shall describe a class of models of quintessential
inflation of this category and look for parameter space which could
comply with  the recent measurement of B mode polarization
spectrum\cite{Ade:2014xna}.

\section{The Model}

The Einstein frame action of the desired model is given by \cite{Wetterich:2013jsa,Hossain:2014xha},
\begin{eqnarray}
&&\mathcal{S} = \int d^4x \sqrt{-g}\bigg[-\frac{\Mpl^2}{2}R+
\frac{k^2(\phi)}{2}\partial^\mu\phi\partial_\mu \phi+V(\phi) \bigg],
~~~
\label{eq:action1}\\
&&k^2(\phi) = \(\frac{\al^2-\t\al^2}{\t\al^2}\)\frac{1}{1+\bet^2
\e^{\al\phi/\Mpl}}+1 \, ,
\end{eqnarray}
where we  assume the form of potential to be exponential,
$V(\phi)=\Mpl^4\e^{-\al\phi/\Mpl}$;
 $\tilde{\alpha}$ controls slow roll such that
$\tilde{\alpha}\gg 1$  and $\beta$ is associated with the scale of
inflation \cite{Wetterich:2013jsa,Hossain:2014xha}.
In the region of large $\phi$, the kinetic function
$k(\phi)\to 1$ which reduces the action to scaling form provided
$\alpha$ is large; nucleosynthesis constraint \cite{Ade:2013zuv} implies that
$\alpha\gtrsim 20$. One also checks that slow roll ensues for small
field limit which continues in the region of large field (see,
discussion below). Hence the kinetic function controls the
inflationary and post inflationary behavior of the field. It is
instructive to have cast the action in non canonical form.

Variation of action (\ref{eq:action1}) with respect to the field
$\phi$ gives its equation of motion, namely,
\begin{equation}
 k^2\Box \phi+k\frac{\partial k}{\partial\phi}\partial^\mu\phi\partial_\mu\phi=
 \frac{\partial V}{\partial\phi} \, .
 \label{eq:eom_phi}
\end{equation}
We can transform the scalar-field part of the action
(\ref{eq:action1}) to its canonical form through the transformation,
$\sigma=\Bbbk(\phi)$ and $k(\phi)=\frac{\partial\Bbbk}{\partial
\phi}$ Thus, (\ref{eq:action1}) becomes
\begin{eqnarray}
 \label{eq:action_E3}
\mathcal{S}_E &=& \int  d^4 x\sqrt{g}\left[-\frac{\Mpl^2}{2}R+\frac{1}{2}
\partial^\mu\sigma\partial_\mu\sigma+V(\Bbbk^{-1}(\sigma))\right].~~~ \,
\end{eqnarray}

In case of small-field approximation, we have $k^2(\phi)\approx
\al^2/\t\al^2$, $\sigma(\phi)\approx
\frac{\alpha}{\tilde\alpha}\phi$ and the potential becomes
$V_\s(\sigma)\approx \Mpl^4 \e^{-\tilde\alpha\sigma/\Mpl}$
\cite{Hossain:2014xha} which can give rise to slow roll for small
values of $\t\al$. Similarly, for very large values of $\phi$, we
have $k^2(\phi)\approx 1$ and
\begin{equation}
\sig\approx\phi-\frac{2}{\t\al}\ln\left(\frac{\bet}{2}\right)+
\frac{2}{\al} \ln\left(\frac{ \t\al\bet}{\al+\t\al}\right)
\end{equation}
and the
potential reads $V_l(\sig)\approx V_{l0}\e^{-\al\sig/\Mpl}$
\cite{Hossain:2014xha}, where $V_{l0}$ is  expressed in terms of
$\Mpl$, $\bet$ and $\t\al$.

\section{Inflation}

The slow roll parameters can be easily expressed in terms of
non-canonical field $\phi$ as
\begin{eqnarray}
\label{eps1}
\epsilon&=&\frac{\Mpl^2}{2}\(\frac{1}{V}\frac{{\rm d}V}{{\rm d}\sig}\)^2
=\frac{\Mpl^2}{2k^2(\phi)}\(\frac{1}{V}\frac{{\rm d}V}{{\rm d}\phi}\)^2
=\frac{\alpha^2}{2k^2(\phi)} ,~~\,\,\\
\eta &=& \frac{\Mpl^2}{V}\frac{{\rm d^2}V}{{\rm d}\sig^2}
= 2\epsilon-\frac{\Mpl}{\alpha}\frac{{\rm d}\ep(\phi)}{{\rm d}\phi}\ ,
\label{eps2}\\
\xi^2&=&\frac{\Mpl^4}{V^2}\frac{{\rm d}V}{{\rm d}\sig}\frac{{\rm d}^3V}{{\rm d}\sig^3}
   = 2\ep\eta-\frac{\al\Mpl}{k^2}\frac{{\rm d}\eta}{{\rm d}\phi} \, .
\label{eps3}
\end{eqnarray}
For $\al\gg1$ and $\t\al\ll1$, we can approximate the slow roll
parameters as,
\begin{equation}
 \ep=\frac{\t\al^2}{2}\left(1+X\right),~~
\eta=\ep+\frac{\t\al^2}{2} ~~{\text{and}}~~ \xi^2=2\t\al^2\ep
\end{equation}
where
$X=\bet^2\e^{\al\phi/\Mpl}$. Power spectra of curvature and tensor
perturbations are
\begin{eqnarray}
 \mathcal{P_R}(k)&=& A_\s(k/k_*)^{n_\s-1+(1/2){\rm d}n_\s/{\rm d}\ln k\ln(k/k_*)}\, , \\
 \mathcal{P}_\rmt(k)&=& A_\rmt(k/k_*)^{n_\rmt} \, ,
\end{eqnarray}
where $A_\s,\; A_\rmt,\; n_\s,\; n_\rmt \;
{\rm and}\; {\rm d}n_\s/{\rm d}\ln k$ are
scalar amplitude, tensor amplitude, scalar spectral index, tensor
spectral index and its running respectively.

Number of e-foldings in the model is given by,
\begin{eqnarray}
 \mathcal{N}\approx \frac{1}{\tilde\alpha^2}\bigg[\ln\left(1+X^{-1}\right)-
 \ln\left(1+\frac{\tilde\alpha^2}{2}\right)\bigg] \, .
 \label{eq:efoldings}
\end{eqnarray}
Considering $\t\al\ll 1$, we can approximate the above expression to
\begin{equation}
 \N\approx \frac{1}{\tilde\alpha^2}\ln\left(1+X^{-1}\right)
\end{equation}
which gives,
\begin{eqnarray}
 \ep(\N)=\frac{\t\al^2}{2} \frac{1}{1-\e^{-\t\al^2\N}} \, .
 \label{eq:epsilon_N}
\end{eqnarray}
For small field approximation  that corresponds to the case
$\t\al^2\gg 1/\N$, we notice that, ($X\ll 1$)
$\ep=\eta/2=\t\al^2/2$. Whereas in  large field limit ($X\gg 1$), we
have $\ep=\eta/2=\t\al^2X/2$ in which case, $\t\al^2\ll 1/\N$. This
means that the transition between the two regions happens when
$\t\al^2\approx 1/\N$.

With the knowledge of $\epsilon(\N,\t\al)$ and $\eta(\N,\t\al)$, the
tensor to scalar ratio ($r$), scalar spectral index ($n_\s$) and its
running (${\rm d}n_\s/{\rm d}\ln k$) read as follows,
\begin{eqnarray}
 r(\mathcal{N},\t\al)&\approx&16\epsilon(\mathcal{N})=\frac{8\t\al^2}{1-\e^{-\tilde\alpha^2\mathcal{N}}} \, ,
 \label{eq:r}\\
 n_\s(\N,\t\al)&\approx& 1-6\ep+2\eta
=1-\tilde{\alpha}^2\coth\(\frac{\t\al^2\N}{2}\)  \, ,
\label{eq:n_s} \\
\frac{{\rm d}n_\s}{{\rm d}\ln k} &\approx&16\ep\eta-24\ep^2-2\xi^2=
-\frac{\t\al^4}{2\sinh^2\(\frac{\t\al^2\N}{2}\)} .~~~~ \label{eq:xi}
\end{eqnarray}

\begin{figure}[h]
\centering
\includegraphics[scale=.6]{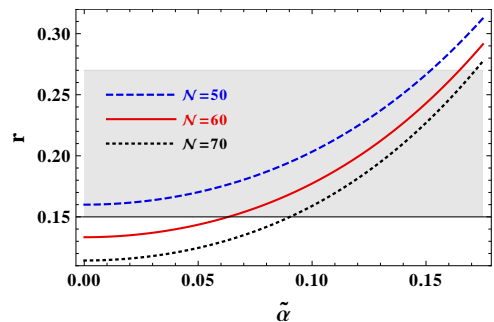}
\caption{The tensor-to-scalar ratio ($r$) versus the model parameter $\t\al$,
for
different e-foldings $\N$. Blue (dashed), red (solid) and black (dotted)
lines correspond to $\N=50$, $60$ and $70$ respectively.
The shaded region marks the BICEP2
constraint on $r$ at $1\sigma$ confidence level, that is
$r=0.2^{+0.07}_{-0.05}$ \cite{Ade:2014xna}.
}
\label{fig:r_alpha}
\end{figure}

In Fig. \ref{fig:r_alpha} we present the variation of the
tensor-to-scalar ratio ($r$) with respect to $\t\al$, for various
numerical values of the number of e-foldings. Additionally, the
shaded region marks the allowed values of $r$ in $1\sig$ confidence
level, given by BICEP2 \cite{Ade:2014xna} collaboration. From BICEP2
results \cite{Ade:2014xna} we have $r=0.2^{+0.07}_{-0.05}$ and Fig.
\ref{fig:r_alpha} clearly shows that
 the values of
$r$ allowed by the BICEP2 can be achieved in the model at hand by
tuning the parameter $\t\al$, for instance $r\approx 0.2$ for
$\t\al=0.12$ and $\N=60$. For these values of $\t\al$ and $\N$,
using Eqs.  (\ref{eq:n_s}) and (\ref{eq:xi}), we find that
$n_\s=0.965$ and ${\rm d}n_\s/{\rm d}\ln k=-0.000522$. While  $n_\s$
satisfies the $1\sig$ bound, $n_\s=0.9603\pm 0.0073$ from {\it
Planck} results \cite{Ade:2013uln}, the value of the running of
$n_\s$ does not satisfy the $1\sig$ bound ${\rm d}n_\s/{\rm d}\ln
k=-0.021^{+0.012}_{-0.010}$ from the same collaboration.

\begin{figure}[h]
\centering
\includegraphics[scale=.42]{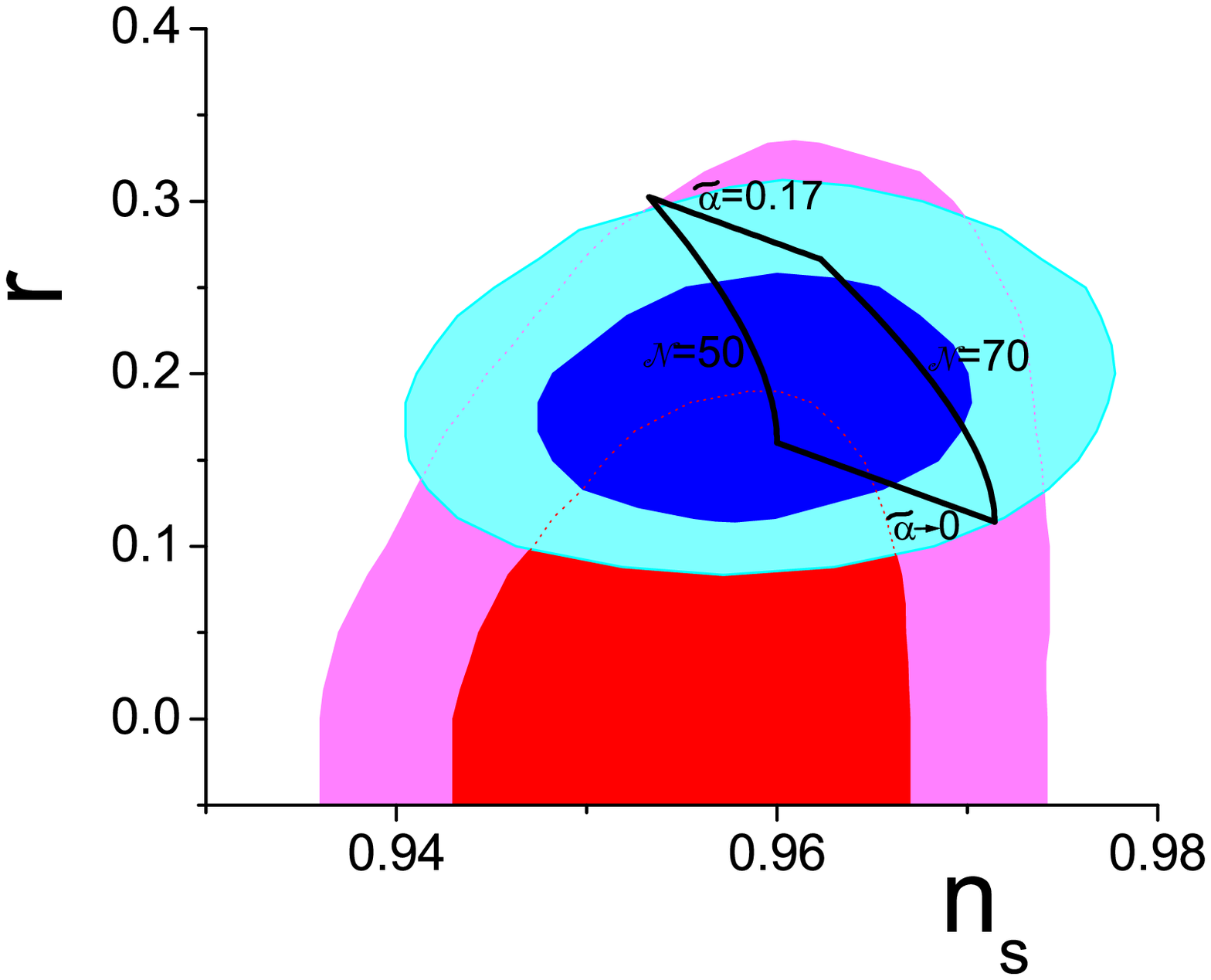}
\caption{1$\sig$ (blue) and 2$\sig$ (cyan) contours for
$Planck+WP+highL+BICEP2$ data, and 1$\sig$ (red) and 2$\sig$
(pink) contours  for
$Planck+WP+highL$ data, on the $n_\s-r$ plane.
The black solid curves bound
the region predicted in our model
for efoldings between $\N=50$ and $\N=70$ and for the parameter $\t\al$
between  $0^+$ and
$0.175$.
The lower  line ($\t\al\to 0$) is for $\N$ from $50$ to 70, the left
curve
($\N=50$) is for $\t\al$ from $0^+$ to 0.17, the right curve ($\N=70$) is
for
$\t\al$ from $0^+$ to 0.17, and the upper line  ($\t\al=0.17$) is for $\N$
from $50$ to 70.
}
\label{fig:ns_r}
\end{figure}

In Fig. \ref{fig:ns_r}, we present the $1\sig$  (blue) and $2\sig$
(cyan) likelihood contours on the $n_\s-r$ plane for the observations
$Planck+WP+highL+BICEP2$ \cite{Ade:2014xna}. For completeness, we
additionally present the $1\sig$ (red) and $2\sig$ (pink) contours,
from the data of $Planck+WP+highL$ \cite{Ade:2013uln}. On top of
these, we depict the predictions of the model at hand. In
particular, the black solid curves bound the region predicted in our
model for efoldings between $\N=50$ and $\N=70$ and for the
parameter $\t\al$ between  $0^+$ and $0.175$. This figure clearly
shows that we can obtain a tensor-to-scalar ratio well within the
$1\sig$ (blue) confidence level by tuning the parameter $\t\al$.
Furthermore, we have $r=-8n_\rmt$, which gives us the range of $n_\rmt$ as
$-0.0338\leq n_\rmt\leq -0.0188$ for the given BICEP2
\cite{Ade:2014xna} range of $r$ in $1\sig$ confidence level.

The COBE normalized value of density perturbations is given by the
following fitting function is \cite{Bunn:1996py},
\begin{equation}
 A_\s=1.91\times
10^{-5}\e^{1.01(1-n_\s)}/\sqrt{1+.75r} \, .
\end{equation}
BICEP2 \cite{Ade:2014xna}
gives the constraint on $r=0.2^{+0.07}_{-0.05}$ and {\it Planck} 2013
results \cite{Ade:2013uln} gives $n_\s=0.9603\pm 0.0073$. So the COBE
normalized value of density perturbations for the best fit values of
$r$ and $n_\s$ taken from the BICEP2 \cite{Ade:2014xna} and {\it
Planck} \cite{Ade:2013uln} observations respectively is
$1.8539\times 10^{-5}$.

On the other hand, the scalar perturbation spectrum is given by
\begin{equation}
 A^2_\s(k)=\frac{V}{\left(150\pi^2\Mpl^4\ep\right)} \, ,
\end{equation}
and at the horizon crossing ($k=k_*=a_*H_*$)
\begin{equation}
 A^2_\s(k_*)=7^{n_{s*}-1}\del^2_H \, .
\end{equation}
 Moreover, the
energy scale of inflation is given by
\begin{eqnarray}
\!\! V_*^{1/4}=\(\frac{7^{n_{s*}-1}r_*}{1-0.07r_*-0.512n_{s*}}\)^{1/4}\!
2.75\times 10^{16} ~\rm GeV \, .
 \label{eq:inf_scale}
\end{eqnarray}
Since in the present model different values of $r$ and
$n_\s$ can be obtained by changing the parameter $\t\al$ and efoldings
$\mathcal{N}$
(see Fig. \ref{fig:r_alpha}), we can use the above formulas in order to
estimate the inflation scale. In particular, for the values of interest,
$r=0.2$ and $n_\s=0.9603$, we get the energy scale of inflation to be
$2.157\times 10^{16}\rm GeV$.

Using COBE normalization, we can also have a
relation between the parameters $\t\al$, $\bet$ and e-foldings $\N$,
\begin{eqnarray}
 \frac{\bet^2\sinh^2\(\t\al^2\N/2\)}{\t\al^2}=6.36\times 10^{-8} \, ,
 \label{eq:al_bet_N}
\end{eqnarray}
which in case of large field approximation (that is for $\t\al^2\N\ll
1$ and $\bet^2\t\al^2=2.5\times 10^{-7}/\N^2$ for the given values)  gives
$r_*=0.2$ and $n_{s*}=0.9603$.

Let us note that the conventional (p)reheating mechanism does not
work in quintessential inflation. However, instant preheating can be
implemented in the scenario under consideration
\cite{Hossain:2014xha}. It is  required that the field potential is
steep in the post inflationary era, allowing radiative regime to
commence. In this case, during scaling regime,
$\Omega_\phi=4/\al^2$; nucleysynthesis from {\it Planck} results
\cite{Ade:2013zuv} then implies that $\al\gtrsim 20$ which in view
of $\tilde{\alpha\ll 1}$
 tells us that inflation ends at a sufficiently
large value of $\phi$ or $X$. Indeed, the large-field slow-roll
regime gives us $\epsilon=\eta=\tilde{\alpha}^2X/2  \ \to\
X_{\rm end}=2/\t\al^2\gg 1$ and the kinetic function
is given by $k^2(\phi)\simeq \alpha^2/(\t\al^2 X)\
\to\ k_{\rm end}\simeq\alpha/\sqrt{2}$.

An important comment regarding the small and large field limit
inflation is in order. As noticed before, the boundary of the two
regions is given by $\t\al=\sqrt{1/\mathcal{N}}$. Thus,
  in case inflation commences in the large field
regime, $\t\al$ needs to be small in order to collect the required
number of e-foldings. However, if inflation begins around the
boundary, the slow roll region is larger and we might improve upon
the numerical values of $\t\al$ for the given number of e-foldings,
thereby giving rise to larger values of $r$. In fact, the large
field approximation does not lead to the desired result in view of
the recent observations.

At the commencement of inflation we have
\begin{equation}
 X_{\rm in}=\frac{1}{\left(\e^{\t\al^2\N}-1\right)} \, ,
\end{equation}
which gives us the value of the
potential at the beginning of inflation
\begin{equation}
 V_{\rm in}=\Mpl^4\bet^2\left(\e^{\t\al^2\N}-1\right) \, .
\end{equation}
Now replacing $\bet$
from   (\ref{eq:al_bet_N}) we acquire
\begin{equation}
 V_{\rm in}=\frac{2.5\times 10^{-7}\t\al^2\Mpl^4}{\left(1-\e^{-\t\al^2\N}\right)} \, .
\end{equation}
$V_{\rm
in}^{1/4}$ gives the scale of inflation as in relation
(\ref{eq:inf_scale}) and for consistency check the value given by
$V_{\rm in}$ should be same as that of  (\ref{eq:inf_scale}). From
Fig. \ref{fig:r_alpha} we can see that $r\approx 0.2$ for
$\t\al\approx 0.12$ and $\N=60$, and for these values of $\al$ and
$\N$ we obtain $V_{\rm in}^{1/4}=2.46\times 10^{16} \rm GeV$, which
matches   the value calculated from Eq. (\ref{eq:inf_scale}) for
$r=0.2$ and $n_\s=0.9603$.

Finally, we also have
\begin{equation}
 \frac{X_{\rm in}}{X_{\rm end}}=\frac{V_{\rm end}}{V_{\rm
in}}= \frac{\t\al^2}{2\left(\e^{\t\al^2\N}-1\right)} \, ,
\end{equation}
which for
large field limit reduces to $V_{\rm end}/V_{\rm
in}=1/\left(2\N\right)$ \cite{Hossain:2014xha}. During the
inflationary phase of the universe the first Friedmann equation
reduces to $3H^2\Mpl^2=V$, and this provides the ratio $H_{\rm
end}/H_{\rm in}$. Therefore, at the end of inflation we have,
\begin{equation}
 H_{\rm end}=\frac{\Mpl\bet\t\al}{\sqrt{6}}=\frac{ 1.02\times
10^{-4}\t\al^2\Mpl}{\sinh\left(\t\al^2\N/2\right)} \, .
\end{equation}

\section{Spectrum of relic gravity waves}

One of the generic
predictions of inflation includes the quantum mechanical production
of relic gravity waves, whose spectral energy density depends upon
the post inflationary equation-of-state parameter $\omega$
\cite{Sahni:1990tx,Sahni:2001qp}. The tensor perturbation is given by,
$\del \g_{ij}=a^2h_{ij}$ where $h_{ij}$ can be represented as
$h_{ij}=h_k(\tau)\e^{-ikx}\e_{ij}$ ($\e_{ij}$ is the polarization
tensor, $\tau$ is the conformal time defined as ${\rm d}\tau={\rm d}t/a$
and k is the comoving wave number defined as $k=2\pi a/\lam$ where $\lam$ 
being the wavelength)
and satisfies the Klein-Gordon equation $ \Box h_{ij}=0$ which reduces
to,
\begin{equation}
 \label{eq:rgw_kg}
 \ddot h_k(\tau)+2\frac{\dot a}
{a}\dot h_k(\tau)+k^2h_k(\tau)=0 \, .
\end{equation}
We will consider power-law expansion of the scale factor $a$ as
$a\sim (t/t_0)^p\sim(\tau/\tau_0)^{(1-2\mu)/2}$ where 
$\mu=(1-3p)/2(1-p)$. For exponential inflation
$\mu=3/2$ and the scale factor goes as $a\sim \tau_0/\tau$,
where $|\tau|<|\tau_0|$. In order to compute the spectral energy
density of gravity waves, we need to compute the Bogoliubov
coefficients (for detailed calculations one can see \cite{Sahni:1990tx,Sahni:2001qp}).
In order to obtain the analytical expression  one
assumes inflation to be exponential. In that case, the spectral
energy density $\tilde{\rho}_\g(k)$, after the transition from de
Sitter to post inflationary phase,  characterized by the
equation-of-state parameter $\omega$, is given by \cite{Sahni:2001qp},
\begin{equation}
\tilde{\rho}_\g(k)\propto k^{1-2|\mu|};~~\mu=\frac{3}{2}
\left(\frac{\omega-1}{3\omega+1}  \right).
\end{equation}

In the model under consideration the post inflationary dynamics is
described by the scalar field in the kinetic regime with
$\omega_\phi=1$, which implies that $\rho_\g \propto k$ thereby a blue spectrum
of gravity wave background. Let us again note that in our case as usual,  $n_\rmt=-r/8$ is small
which we ignored when assumed inflation to be exactly exponential.
The blue spectrum in our proposal is solely attributed to the kinetic regime that
follows quintessential inflation irrespective of the underlying model.

As demonstrated
in Refs. \cite{Sahni:2001qp,Giovannini:1998bp,Giovannini:1999bh,Giovannini:1999qj},
the gravitational waves amplitude
increases during the kinetic regime and this might come into
conflict with nucleosynthesis at the commencement of radiative
regime, in case the kinetic regime is long. Since the standard
mechanism does not work here, One assumes that radiation with energy
density $\rho_\r$ is produced via an alternative mechanism. The ratio
of the field energy density to $\rho_\r$, can be compared with the
ratio of the energy density in gravity waves $\rho_\g$ to radiation
energy density at the beginning of radiative era, namely,
\begin{equation}
\(\frac{\rho_\phi}{\rho_{\rm r}}\)_{\rm end}=\frac{3\pi}{64 h^2_{\rm GW}}
\(\frac{\rho_\g}{\rho_\r}\)_{\rm eq},
\end{equation}
where,
\begin{equation}
 h^2_{\rm GW}=\frac{H_{\rm in}^2}{8\pi\Mpl^2}
 =\frac{3.315\times 10^{-9}\t\al^2}{1-\e^{-\t\al^2\N}} \, ,
 \label{eq:hgw}
\end{equation}
is the square of dimensionless gravity wave amplitude. Nucleosynthesis
imposes  a stringent constraint on the ratio of energy densities at
equality, that is $\rho_\g/\rho_\r\lesssim 0.01$, thereby giving rise to the
lower bound of $\rho_\phi/\rho_\r$ at the end of inflation which reads
\begin{equation}
 \rho_{\r,\rm end}\geq \frac{3.517\times 10^{-14}\Mpl^4\t\al^6
\e^{\t\al^2\N/2}}{\sinh^3\left(\t\al^2\N/2\right)} \, ,
\end{equation}
and it also gives
$T_{\rm end}=\rho^{1/4}_{r,\rm end}$. Now the bound on $r$ from BICEP2
\cite{Ade:2014xna} gives the bound on $\t\al$ as $0.063\leq\t\al\leq 1.83$
for $\N=60$. For $\t\al=0.12$ and $\N=60$, $r\approx 0.2$ and 
we get the bound on the temperature 
at the end of inflation as, 
$T_{\rm end}\geq 6.65\times 10^{13}\rm GeV$. This
condition cannot be met for instance if reheating is attempted via
gravitational particle production, which is an inefficient mechanism.

It is possible to circumvent the problem of over-production of
relic gravity waves  if an efficient mechanism such
as instant preheating \cite{Felder:1998vq,Felder:1999pv,Campos:2004nc}
is implemented \cite{Sahni:2001qp,Hossain:2014xha}.

The spectral
energy density parameter of the relic gravitational
wave is defined as,
\begin{equation}
 \Omega_{\rm GW}(k)=\frac{\t\rho_\g(k)}{\rho_{\rm c}} \, ,
\end{equation}
where $\rho_{\rm c}$ is the critical energy density and
(detailed calculations one can see Ref. \cite{Sahni:2001qp})
\begin{eqnarray}
&&\Omega_{\rm GW}^{\rm (MD)}= \frac{3}{8\pi^3}h_{\rm GW}^2
\Omega_{\m 0}\(\frac{\lam}{\lam_\h}\)^2 \, ,  \lam_{\rm MD}<\lam\leq \lam_\h  ,~~~~~~\\
&&\Om_{\rm GW}^{\rm (RD)}(\lam)=\frac{1}{6\pi}h_{\rm GW}^2\Omega_{\r 0} \, ,
~~~~~~~~~ \lam_{\rm RD}<\lam\leq\lam_{\rm MD} \, ,~~~~~ \\
&&\Om_{\rm GW}^{\rm (kin)}(\lam)=\Om_{\rm GW}^{\rm (RD)}\(\frac{\lam_{\rm RD}}{\lam}\) \, ,
~~~~~  \lam_{\rm kin}<\lam\leq\lam_{\rm RD} \, ,~~~~~
\end{eqnarray}
where,
\begin{eqnarray}
 \lam_\h &=& 2cH_0^{-1}\, , \\
 \lam_{\rm MD} &=& \frac{2\pi}{3}\lam_\h\(\frac{\Om_{\r 0}}{\Om_{\m 0}}\)^{1/2} \, , \\
 \lam_{\rm RD}&=& 4\lam_\h\(\frac{\Om_{\m 0}}{\Om_{\r 0}}\)^{1/2}\frac{T_{\rm MD}}{T_{\rm rh}} \, ,\\
 \lam_{\rm kin} &=& cH_{\rm kin}^{-1}\(\frac{T_{\rm rh}}{T_0}\)\(\frac{H_{\rm kin}}{H_{\rm rh}}\)^{1/3} \, ,
\end{eqnarray}
where matter, radiation and kinetic energy dominated epochs
are represented by ``MD'', ``RD'' and ``kin'' respectively. $H_0$, $\Om_{\m 0}$ and
$\Om_{\r 0}$ are the present values of Hubble parameter,
matter and radiation energy density parameters respectively.
reheating temperature and Hubble parameter are represented by
$T_{\rm rh}$ and $H_{\rm rh}$ respectively and we have taken
reheating temperature and Hubble parameter approximately same as
the temperature and Hubble parameter at the end of inflation.

\begin{figure}[h]
\centering
\includegraphics[scale=.64]{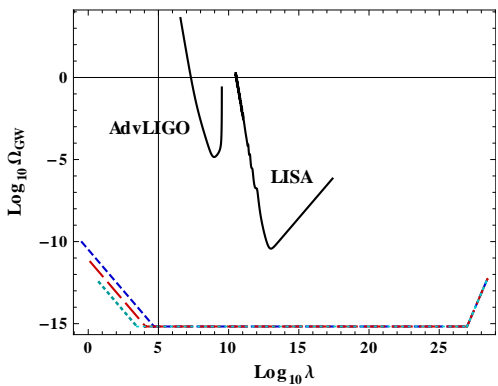}
\caption{The spectral energy density of the relic gravity wave background
versus the wavelength $\lam$. Blue (small dashed), red (long
dashed) and cyan
(dotted) lines correspond to reheating temperature $7\times 10^{13}\rm GeV$,
$2.5\times 10^{14}\rm GeV$ and $8\times 10^{14}\rm GeV$ respectively. We have taken
$\t\al=0.12$ and $\N=60$. Black solid lines represent the sensitivity curves of
advanced LIGO and LISA.
}
\label{fig:RGW}
\end{figure}

\begin{figure}[h]
\centering
\includegraphics[scale=.64]{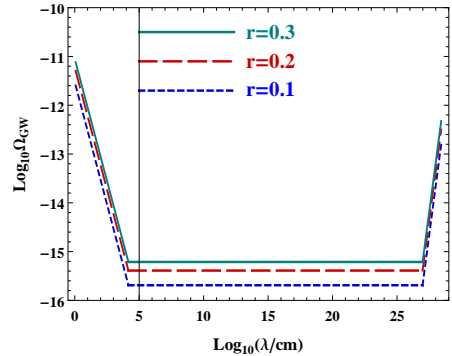}
\caption{The spectral energy density of the relic gravity wave background
versus the wavelength $\lam$. Blue (small dashed), red (long dashed) and cyan
(solid) lines correspond to the tensor to scalar ratio $r=0.1,\; 0.2\; {\rm and}\; 0.3$
respectively with reheating temperature $10^{14}\rm GeV$. We have taken
$\t\al=0.12$ and $\N=60$.}
\label{fig:RGW_r}
\end{figure}

Fig. \ref{fig:RGW} shows the spectrum of the spectral energy
density of relic gravitational waves with   wavelength
$\lam$. Sensitivity curves of
advanced LIGO \cite{aLIGO} and LISA \cite{LISA1} are also
depicted. Additionally, in Fig. \ref{fig:RGW_r} we present the spectrum
of relic gravitational waves for different tensor to scalar ratio. We can
write the amplitude of the relic gravitational wave ($h_{\rm GW}$) in terms
of the tensor-to-scalar ratio ($r$) by using Eq. (\ref{eq:r}) and Eq.
(\ref{eq:hgw}), and it is
given as $h_{\rm GW}^2=3.315\times 10^{-9}r/8$. This implies that the 
square of the amplitude is directly proportional to $r$, and since the 
spectral energy density parameter of the relic gravitational wave ($\Om_{\rm GW}$) is 
proportional to the square of the amplitude, $\Om_{\rm GW}$ also gets
increased 
with increasing $r$. This effect can be seen in Fig. \ref{fig:RGW_r}.

\section{Late time Evolution}

Finally, let us point out for completeness that the late time acceleration,
although not the subject here, can be achieved
in the scenario by including massive neutrino matter, non-minimally
coupled to the field \cite{Wetterich:2013jsa,Hossain:2014xha}.
At late stage, when neutrinos become non
relativistic, the field potential, originally a steep run-away
potential, acquires a minimum allowing the exit from the scaling
regime to dark energy \cite{Hossain:2014xha}.

\section{Conclusions}

In this paper we have investigated a class of models that can
successfully give rise to quintessential inflation. The Lagrangian
of the single field system under consideration contains three free
parameters $\tilde{\alpha}$, $\alpha$ and $\beta$ such that $\beta$
is related to scale of inflation and $\tilde{\alpha}$ defines the
tensor to scalar ratio $r$ for a given number of efolds. As for $\alpha$, it is fixed
by the post inflationary requirements, namely, nucleosynthesis
constraint \cite{Ade:2013zuv}.

For the observed values of $r$ from BICEP2 \cite{Ade:2014xna} 
and $\mathcal{N}=60$, the parameter 
$\tilde{\alpha}$ in the model ranges from $0.063$ to $0.183$
consistent with the BICEP2 measurements,
(see Fig. \ref{fig:r_alpha}) such that the scale of inflation in
this case is around the GUT scale. The distinguished feature of the
model includes a blue spectrum of stochastic background of relic
gravitational waves produced during inflation. The blue spectrum of
relic gravity waves associated with the kinetic regime after
inflation, is a generic feature of quintessential inflation
irrespective of an underlying model \cite{Sahni:2001qp,Sami:2004xk,
Giovannini:1998bp,Giovannini:1999bh,Giovannini:1999qj}. However,  the amplitude of
relic gravity waves naturally depends upon the tensor to scalar
ratio of perturbations and we have quoted here $\Omega_{\rm GW}$ in
accordance with the observed values of $r$.
 Fig. \ref{fig:RGW_r} 
shows the  $r$ dependence of spectral energy density 
parameter ($\Om_{\rm GW}$). 
We should emphasize that
we have  neglected $n_\rmt$, the tilt of inflationary spectrum, in order to
felicitate the analytical calculation. We reiterate that the blue
spectrum here is nothing to do with blue tilt seen in BICEP2; the
former is the consequence of kinetic regime which is a general
feature in scenarios of quintessential inflation. We should also
negligibly small values of running of the spectral index.
 note that the scenario under consideration predicts

The BICEP2 findings, if confirmed, would rule out a large number of
models including the currently favourite Starobinsky model. In purely
theoretical perspective, the GUT scale of inflation as envisaged by
the said measurements would throw a big challenge to model building
in the framework of effective field theories. We hope that the
forthcoming announcement from {\it Planck} collaboration and future
observations would clarify the related issues and the same is
eagerly awaited.

\section{Acknowledgments}

We thank S.G. Ghosh, V. Sahni and T. Souradeep for useful discussions.
MWH acknowledges CSIR, Govt. of India for financial support through
SRF scheme. The research of ENS is implemented within the framework of the
Action ``Supporting Postdoctoral Researchers'' of the Operational Program
``Education and Lifelong Learning'' (Actions Beneficiary: General Secretariat
for Research and Technology), and is co-financed by the European Social Fund
(ESF) and the Greek State.

\end{document}